\documentclass[aps,prx,reprint,showpacs]{revtex4-1}
\usepackage{braket}
\usepackage{graphicx}
\usepackage{dcolumn}
\usepackage{bm}
\usepackage{epsfig}
\usepackage{amsmath}
\usepackage[usenames,dvipsnames]{color}
\usepackage{booktabs}
\begin{document}

\title{Complex superconductivity in the noncentrosymmetric compound Re$_{6}$Zr}

\author{Mojammel A. Khan}
\email{mkhan19@lsu.edu}
\affiliation{Department of Physics and Astronomy, Louisiana State University, Baton Rouge, LA 70808}

\author{A. B. Karki}

\affiliation{Department of Physics and Astronomy, Louisiana State University,  Baton Rouge, LA 70808}

\author{T. Samanta}
\affiliation{Department of Physics and Astronomy, Louisiana State University,  Baton Rouge, LA 70808}

\author{D. Browne
}
\affiliation{Department of Physics and Astronomy, Louisiana State University,  Baton Rouge, LA 70808}
\author{S. Stadler}

\affiliation{Department of Physics and Astronomy, Louisiana State University,  Baton Rouge, LA 70808}
\author{I. Vekhter}

\affiliation{Department of Physics and Astronomy, Louisiana State University,  Baton Rouge, LA 70808}

\author{Abhishek Pandey}
\email{Current address: Department of Physics and Astronomy, Texas
A\&M University, College Station, Texas 77840-4242, USA.}
\affiliation{Department of Physics and Astronomy, Louisiana State University,  Baton Rouge, LA 70808}

\author{P. W. Adams}

\affiliation{Department of Physics and Astronomy, Louisiana State University,  Baton Rouge, LA 70808}

\author{S. Teknowijoyo}
\affiliation{Ames Laboratory and Department of Physics and Astronomy, Iowa State University, IA 50011}

\author{K. Cho}
\affiliation{Ames Laboratory and Department of Physics and Astronomy, Iowa State University, IA 50011}

\author{R. Prozorov}
\affiliation{Ames Laboratory and Department of Physics and Astronomy, Iowa State University, IA 50011}

\author{D. E. Graf}
\affiliation{National High Magnetic Field Laboratory, Tallahassee, FL 32310}

\author{D. P. Young}
\email{dyoung@phys.lsu.edu}
\affiliation{Department of Physics and Astronomy, Louisiana State University, Baton Rouge, LA 70808}

\date{\today}
\begin{abstract}
We report the electronic structure, synthesis, and measurements of the magnetic, transport, and thermal properties of the polycrystalline noncentrosymmetric compound Re$_{6}$Zr. We observed a bulk superconducting transition at temperature $T_{c}$ $\sim$ 6.7 K, and measured the resistivity, heat capacity, thermal conductivity, and the London penetration depth below the transition, as well as performed doping and pressure studies. From these measurements we extracted the critical field, and the superconducting parameters of Re$_{6}$Zr. Our measurements indicate a relatively weak to moderate contribution from a triplet component to the order parameter, and favor a full superconducting gap, although we cannot exclude the existence of point nodes based on our data.

\end{abstract}


\maketitle

\section{Introduction}
The topic of superconductivity in systems lacking spatial inversion symmetry has undergone a resurgence since the discovery of superconductivity in CePt$_{3}$Si~\cite{book}. In a material lacking an inversion center, parity, and hence spin, is no longer a good quantum number. Consequently, the conventional classification of the superconducting  states into a spin singlet with the spatially symmetric ($s$,$d$-wave), or triplet with antisymmetric ($p$,$f$-wave) wave functions of the Cooper pairs no longer applies. Instead, in a noncentrosymmetric superconductor (NCS), the order parameters possessing different spatial symmetry properties (transforming according to different representations of the point group of the crystal) generally have mixed singlet and triplet character. Technically, lack of inversion symmetry dictates that the band spin-orbit (SO) coupling is antisymmetric in the crystal momentum, ${\bm k}$, leading to energy splitting of the Bloch states at the same ${\bm k}$ with opposite spins. In most common situations the strength of this antisymmetric spin-orbit coupling (ASOC) significantly exceeds the superconducting energy gap, pre-empting pairing between ASOC-split bands, resulting in mixed singlet-triplet states~\cite*{sing,rashba,rashba2}. If the triplet component is significant, the superconducting gap may become highly anisotropic, and develop line or point nodes~\cite*{prp1,prp2,prp3,prp4}. It is important to note that there is no symmetry requirement for gap anisotropy in NCSs. Strong anisotropy is rare in such materials~\cite*{weakly,bauerncs2,trs3}, and most frequently appears from the projection of the pairing interaction onto the ASOC-split bands, as, for example, in Li$_{2}$Pt$_{3}$B~\cite*{LiPtB}.

Very recently, muon spin relaxation ($\mu$SR) measurements on two NCSs,  LaNiC$_{2}$ \cite{lanictrs} and Re$_{6}$Zr \cite{sing}, indicated time reversal symmetry (TRS) breaking in both the systems by detecting the apearance of spontaneous magnetic field at the onset of superconductivity. So far only a handful of unconventional superconductors, e.g. Sr$_{2}$RuO$_{4}$~\cite*{trs,trs2}, UPt$_{3}$, (U,Th)Be$_{13}$~\cite*{u,u2}, (Pr,La)(Os,Ru)$_{4}$Sb$_{12}$~\cite*{la}, PrPt$_{4}$Ge$_{12}$, and LaNiGa$_{2}$~\cite*{ge,lani} were found to exhibit this property. LaNiC$_{2}$ and Re$_{6}$Zr are the only two NCSs so far that show evidence for the TRS-broken states. Generally, since the ASOC has the full symmetry of the lattice, absence of inversion symmetry by itself cannot lead to the TRS breaking: strong electron-electron interactions and/or unconventional pairing mechanism are required to stabilize such a state. Indeed, it was suggested that a purely-triplet state emerges in LaNiC$_{2}$ due to electron-electron correlations allowing the TRS breaking\cite{lanictrs}. Similarly, if Re$_{6}$Zr supports the TRS state, as is indicated by the $\mu$SR measurements~\cite{sing}, it must be due to an unconventional pairing mechanism. For such a mechanism the gap anisotropy is required by symmetry, and the pairing states suggested in Ref.~\cite{sing} indeed possess either line or point nodes. While at present there is no other corroborating evidance available for TRS breaking, we invesitagte the possible consequences of it on the superconducting properties. Even though superconductivity in Re$_{6}$Zr was first reported in 1961~\cite{matthias}, there have been no comprehensive studies of the electronic structure and physical properties of Re$_{6}$Zr at low temperature.

In this paper, we present the synthesis, characterization, electronic band structure, resistivity, heat capacity, thermal conductivity, and low-temperature penetration depth of Re$_{6}$Zr, along with results from chemical doping and a physical hydrostatic pressure study. Thermal and penetration depth measurements are very useful in providing information on the nature of the pairing state and the electron-electron interactions.  In conventional BCS-like superconductors, the thermal conductivity, $\kappa$$(T)$, decreases below $T_{c}$~\cite*{bardeen}.  However, in unconventional superconductors (non BCS-type), such as strongly correlated systems, high $T_{c}$ cuprates, and iron pnictides, the thermal conductivity often increases upon cooling below $T_{c}$~\cite*{ho,lee,spin,iron,iron2,zhang}.  For example, it was suggested that over-doped samples in Co-doped BaFe$_2$As$_2$ show an enhancement in the electronic contribution to their thermal conductivity below $T_c$, due to an increase in the quasiparticle mean free path and the presence of a nodal gap structure~\cite*{reid}.  More recent measurements on the same system attribute the enhancement in thermal conductivity to the formation of a spin gap with a reduction in electron scattering from magnetic spin fluctuations~\cite*{spin}. At low temperatures, $T\ll T_c$, the behavior of the electronic thermal conductivity, heat capacity, and the penetration depth carry information about the nodal structure of the gap.

The structure of Re$_{6}$Zr is $\alpha$-Mn, cubic with space group I$\bar4$3m. The unit cell has 58 atoms that occupy four distinct crystallographic  sites. A recent study~\cite*{sing} showed bulk Re$_{6}$Zr has a superconducting transition of 7 K. From measurements performed in this work, several superconducting parameters, such as the coherence length ($\xi$), penetration depth ($\lambda$), and the upper critical field ($H_{c2}$) were estimated. $H_{c2}$ is important for NCS superconductors, since its value in comparison with the Pauli limiting field can suggest the existence of a triplet component to the pairing.  We also performed specific heat measurements, which verified a bulk superconducting transition. Thermal conductivity showed an enhancement in the electronic contribution below $T_c$, very similar to that observed in Co-doped BaFe$_2$As$_2$. The enhancement in the thermal conductivity is suppressed with the application of an external magnetic field. Our results at low temperature do not give evidence for the existence of nodal quasiparticles, and are most consistent with a fully gapped superconductor. While it is possible that the contribution of the excitations from linear point nodes is sufficiently small to be compatible with the data, we do not find good evidence for line nodes. This severely restricts the possible order parameters in Re$_6$Zr.

\section{Experiment}

Polycrystalline samples of Re$_{6}$Zr were made by arc melting stoichiometric amounts of pure Zr slug (99.99\% Alfa Aesar) and Re slug (99.99\% Alfa Aesar) under a partial pressure of ultra high purity argon gas on a water-cooled copper hearth with a tungsten electrode. The button of Re$_{6}$Zr was flipped several times and remelted to ensure a homogeneous sample. Mass loss during the synthesis was negligible, and the sample formed a uniform and hard button. Several off-stoichiometric samples were made to check the effect of stoichiometry on $T_c$. We also synthesized and studied samples doped with Hf, Ti, W, and Os to check the effect of doping on $T_c$.

The crystal structure and phase purity of the arc melted samples were investigated by powder X-ray diffraction using a small portion of powdered sample on a PANalytical Empyrean multi-stage X-ray diffractometer with Cu K$\alpha$ radiation ($\lambda$ = 1.54 \AA). The system has a $\theta$-2$\theta$ geometry, and data were taken from $10^{\circ}$ to $90^{\circ}$ at a constant scan of $2^{\circ}$ per minute at room temperature. Elemental composition was determined with a JSM-6610LV high performance scanning electron microscope (SEM) equipped with an energy-dispersive spectrometer (EDS).

The electrical resistivity was measured using a standard four-probe ac technique at 27 Hz with an excitation current of 1-3 mA, in which small diameter Pt wires were attached to the sample using a conductive epoxy (Epotek H20E).  Data were collected beween 1.8 to 290 K and in magnetic fields up to 9 T using a Quantum Design, Physical Property Measurement System (PPMS). The specific heat was measured in the PPMS using a time-relaxation method between 2 and 20 K at 0 and 9 T.  Magnetic susceptibility was also measured in the PPMS in a constant magnetic field of 30 Oe; the sample was zero-field-cooled (ZFC) to 1.8 K, and then magnetic field was applied, followed by heating to 10 K and then cooled down again to 1.8 K [field-cooled (FC)]. The low temperature upper critical field was measured in a 35 T resistive magnet at the National High Magnetic Field Laboratory (NHFML) using a four-probe ac technique with a 3 mA excitation current. The thermal conductivity was measured with the PPMS's thermal transport option, which uses a low-frequency square-wave heat pulse to create a temperature gradient across the sample.  The temperature dependence of the superconducting penetration depth was measured in a $^{3}$He fridge with a 9 T magnet at Ames Laboratory using a tunnel-diode resonator (TDR) oscillating at 14 MHz and at temperatures down to 0.5 K. Further details of the measurement and calibration can be found elsewhere\cite{depth1,depth2,depth3}.

Measurement of the transition temperature under applied hydrostatic pressure ($P$) was carried out in a commercial BeCu cylindrical pressure cell (Quantum Design) within a Magnetic Properties Measurement System (Quantum Design, MPMS SQUID magnetometer). Daphne 7373 oil was used as the pressure-transmitting medium. The value of the applied pressure was calibrated by measuring the shift of the superconducting transition temperature of Pb, which was used as a reference manometer ($T_{c}$ of Pb is $\sim$ 7.19 K at ambient pressure)\cite{pressure}.

\section{result \& discussion}

\subsection{Characterization}

\begin{figure}[h!]
\includegraphics[scale=0.45]{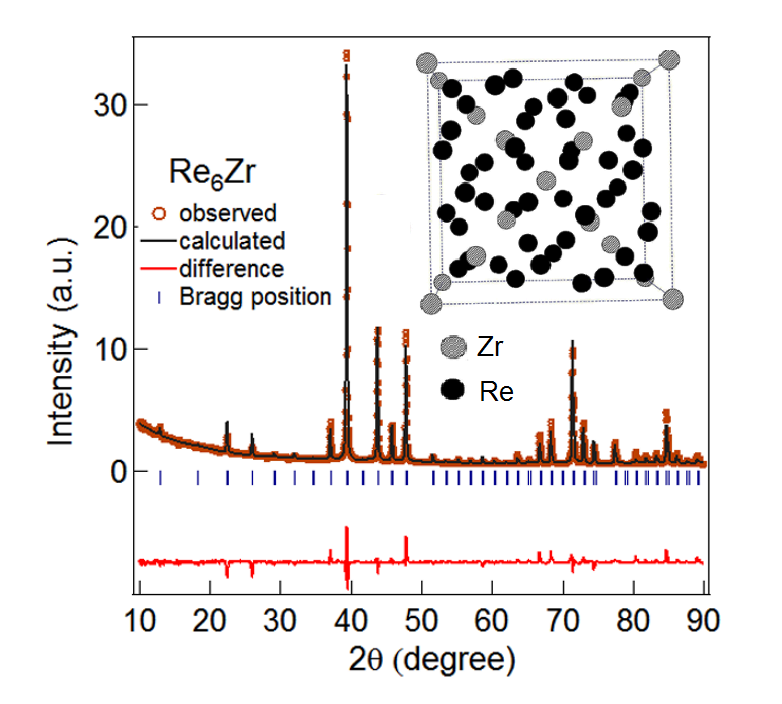}
\caption{Powder XRD pattern (open circles) of Re$_{6}$Zr at room temperature. The solid line represents the Rietveld refinement
fit calculated for the $\alpha$-Mn cubic-type structure
with space group I$\bar4$3m. The inset shows a schematic view of the crystal structure of Re$_{6}$Zr.}\label{xray}
\end{figure}

The XRD pattern of polycrystalline Re$_{6}$Zr is shown in Fig.~\ref{xray}. Rietvelt refinement of the XRD data indicates the sample was single phase with a cubic cell parameter of $a$ = 9.6989 $\pm$ 0.0002\AA, which is in good agreement with published data~\cite*{sing}. A schematic view of the crystal structure is shown in the inset of Fig.1. Re$_{6}$Zr forms in the $\alpha$-Mn cubic structure with the primitive Bravais lattice I$\bar4$3m (space group 217). This particular structure lacks a center of inversion. SEM data, utilizing EDS, confirmed the atomic ratio is approximately 6:1.

The electronic band structure of Re$_6$Zr was calculated using the WIEN2K full-potential linearized augmented plane wave(LAPW) software package~\cite*{wien} using the Generalized Gradient Approximation
exchange-correlation potential~\cite*{lapw}. The room temperature lattice constant of 9.6989 \AA $\space$ was used and the cutoff in the LAPW basis was varied from $RK$$_{\rm max}$ = 7.00 to $RK$$_{\rm max}$ = 8.00 to ensure convergence. A 19 $\times$ 19 $\times$ 19 mesh of points was used for the Brilluoin zone integration that employed the modified tetrahedron method. One set of calculations was done omitting the SO interaction for the valence bands. A second set of self-consistent calculations included the SO interactions for the valence bands using a scalar relativistic approximation.

\begin{figure}[h!]
\includegraphics[scale=0.75]{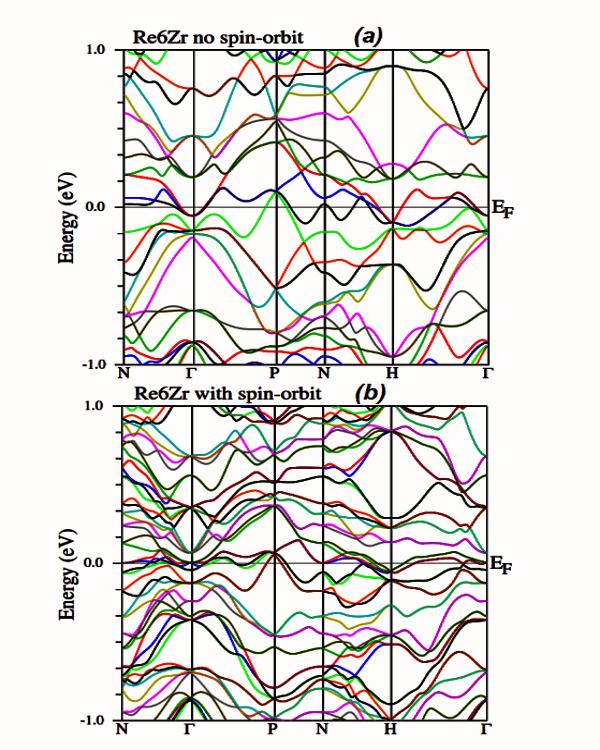}
\caption{Section of the calculated electronic band structure of Re$_{6}$Zr along high symmetry directions within the range of $\pm$ 1 eV around $E_F$. Panel (a) shows bands without SO coupling and (b) shows bands with SO coupling.}\label{band}
\end{figure}

\begin{figure}[h!]
\includegraphics[scale=0.48]{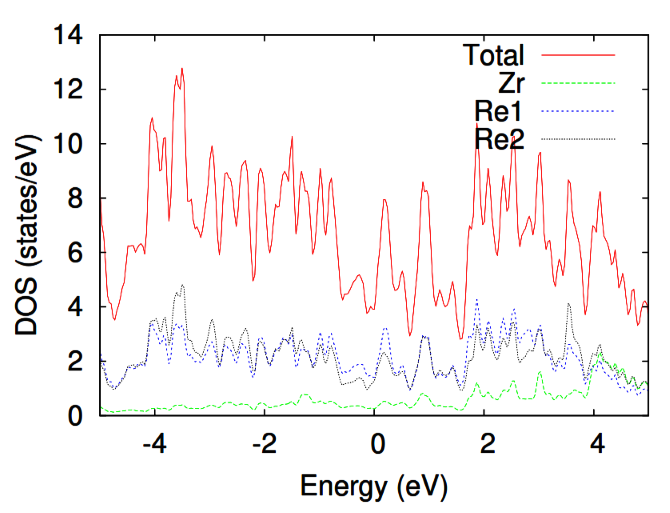}
\caption{Section of the total and individual atom projected DOS (in units of states e$V^{-1}$) of Re$_{6}$Zr within the $\pm$4 eV energy range around $E_F$. Result here is shown without SO coupling.}\label{dos}
\end{figure}

The results of these calculations are presented in Fig.~\ref{band} for the band structure and Fig.~\ref{dos} for the density of states (DOS). The DOS near the Fermi level is almost entirely composed of Re and Zr $d$ bands, however, since the ratio of Re to Zr is 6:1, Re-$d$ bands comprise the majority of the states. There are 4 bands at $\Gamma$ about 0.2 eV below the Fermi surface, as seen in Fig.~\ref{band}(a), and two of them look like a Dirac point. SO coupling lifts the bands very close to the Fermi level and  splits the spin degeneracy which can be seen in Fig.~\ref{band}(b). The band splitting due to the SO interaction is about 30 meV and is comparable to that of Li$_2$Pd$_3$B~\cite*{book}. Its worthwhile to note that the band splitting due to ASOC in Li$_2$Pt$_3$B is about 200 meV, and it has an anisotropic superconducting gap~\cite*{LiPtB,book}, while Li$_2$Pd$_3$B has a fully gapped isotropic order parameter.

\begin{figure}[h!]
\includegraphics[scale=0.45]{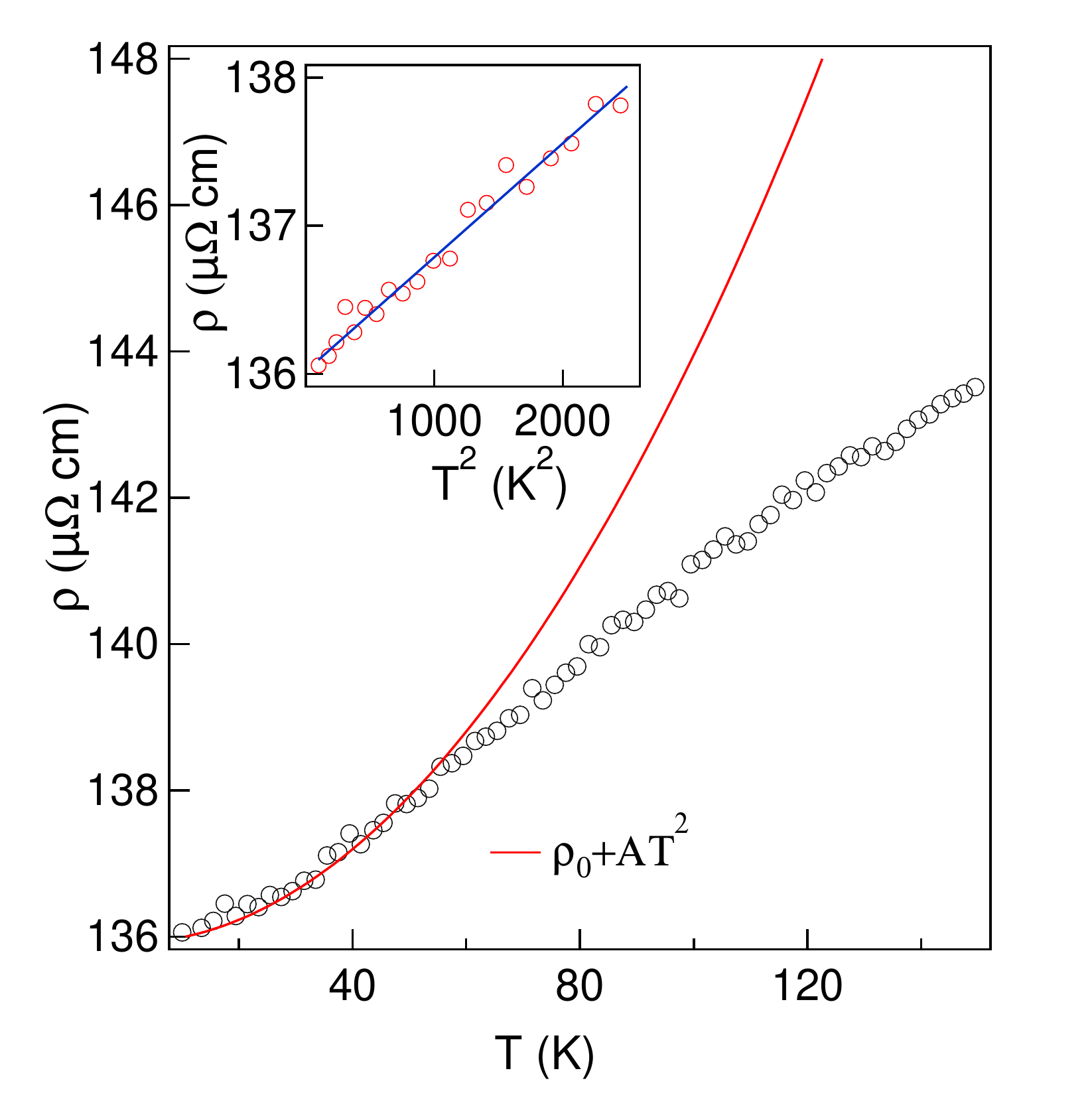}
\caption{Normal state temperature dependence of the electrical resistivity of Re$_{6}$Zr. The solid line is a fit to the data as discussed in the text. Inset shows the data between 10-50 K and blue solid line indicates a line fit to the data.}\label{resisivitynormal}
\end{figure}

The normal state temperature dependence of the resistivity of Re$_{6}$Zr between 10 K and 150 K is shown in Fig.~\ref{resisivitynormal}. The resistivity is metallic, and an inflection point in $\rho(T)$ at $\approx$ 50 K is observed. The normal state resistivity ($\approx$ 150 $\mu$$\Omega$ cm at room temperature) is typical of polycrystalline metallic materials. The residual resistivity ratio (RRR), $\rho_{290\rm K}$/$\rho_{10\rm K}$ = 1.10, is small, which suggests the transport in the sample is dominated by disorder. The low temperature resistivity data were fitted as shown in the inset of Fig.~\ref{resisivitynormal} to the power law,
\begin{equation}
\rho = \rho_0 + AT^\alpha.
\end{equation}

Here, $\alpha$ = 2, the residual resistivity $\rho_{0}$ $\approx$ 135 $\mu$$\Omega$ cm, and the coefficient $A$ = 0.0079 $\pm$ 0.0002 $\mu$$\Omega$ cm/K$^{2}$. The fit describes the data reasonably well between 10 and 50 K, suggesting a Fermi-liquid like temperature dependence at low temperature in the normal state.

\begin{figure}[h!]
\includegraphics[scale=0.50]{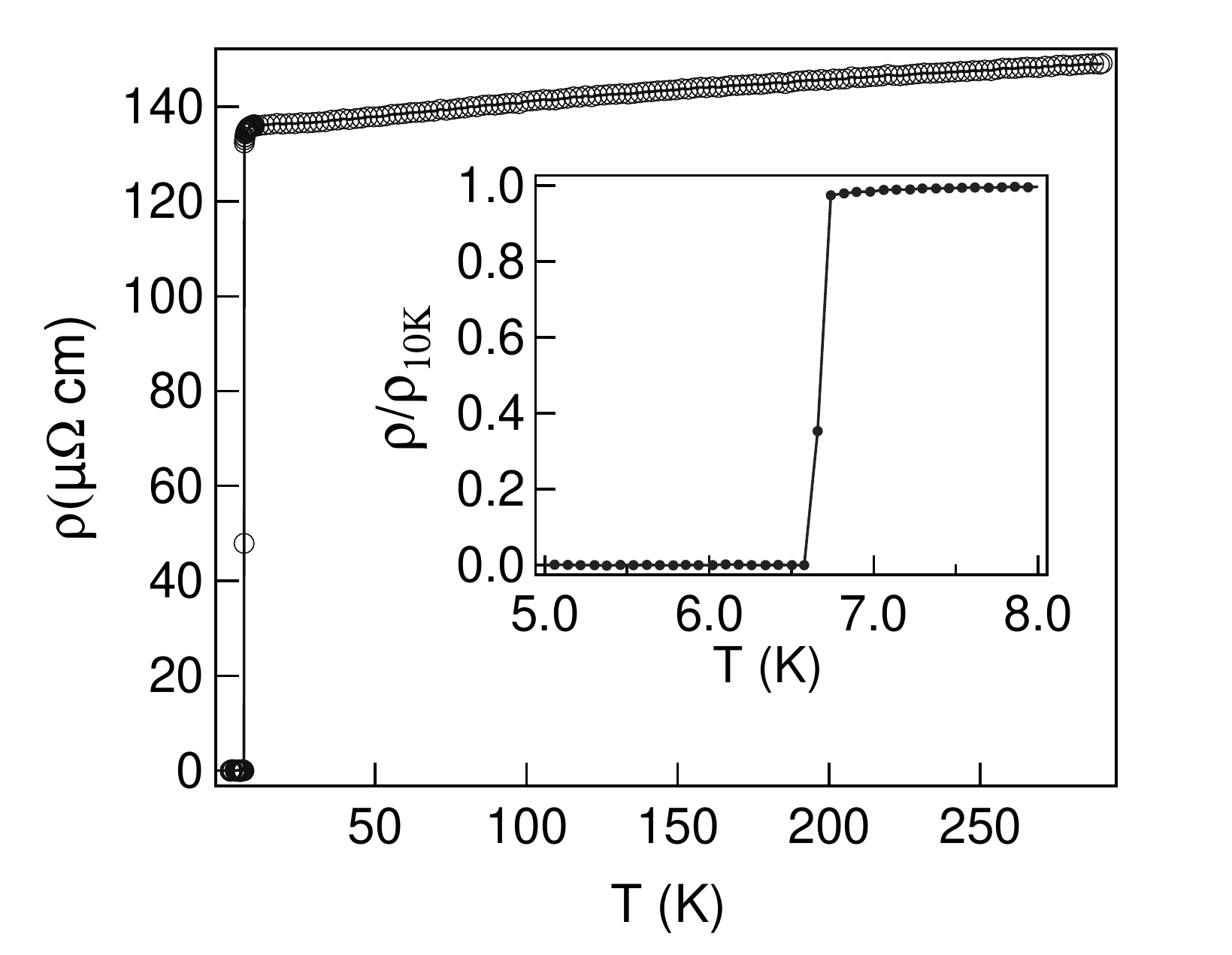}
\caption{Temperature dependence of the electrical resistivity of Re$_{6}$Zr from 2 K to 290 K with the inset showing a sharp superconducting transition at 6.7 K.}\label{tc}
\end{figure}

The value of $\rho_{0}$ is large, most likely due to the polycrystalline nature of the sample. The Kadowaki-Woods ratio~\cite{Kadowaki2} (KWR) $A$/$\gamma$$^{2}$, where $\gamma$ is the sommerfeld coefficient, is taken as a measure of the degree of electron correlations in the material. From the heat capacity data $\gamma$ was estimated as 27.5 mJ mol$^{-1}$K$^{-2}$, and the KWR was found to be $A$/$\gamma$$^{2}$ $\approx$ 10.44 $\mu$$\Omega$$\space$ cm$\space$ mol$^{2}$$\space$ K$^{2}$$\space$J$^{-2}$. This KWR value is typical of heavy fermions and suggests Re$_{6}$Zr is a strongly correlated electron system\cite{Kadowaki2,KADOWAKI}. The temperature dependent resistivity from 2 to 290 K is shown in Fig.~\ref{tc}. A sharp superconducting transition is observed at 6.7 K (inset Fig.~\ref{tc}). The 90\% to 10\% transition width is less than 0.05 K.

\begin{figure}[h!]
\includegraphics[scale=0.45]{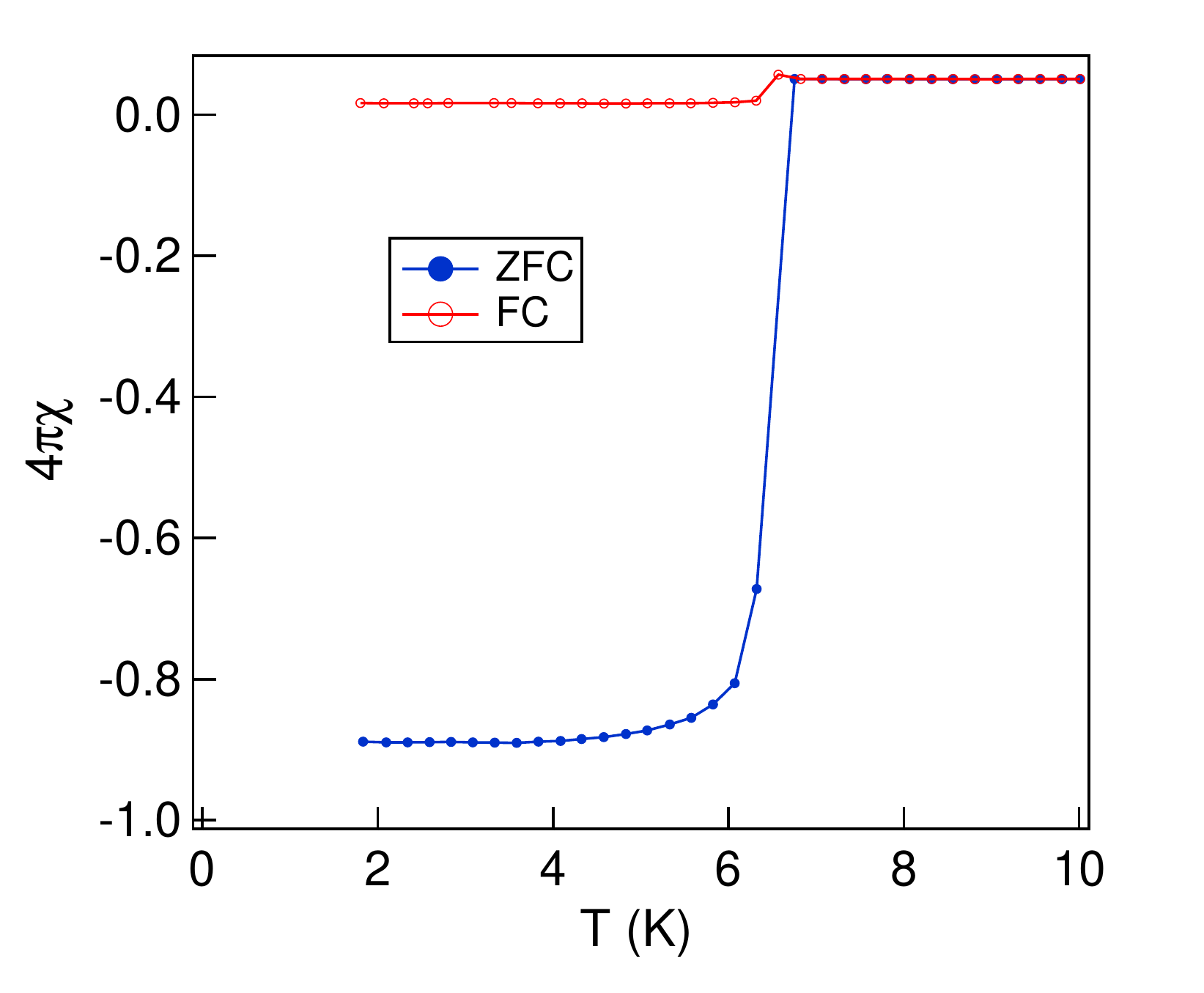}
\caption{ ZFC and FC susceptibilities for Re$_{6}$Zr at a constant field of 30 Oe. The value of the ZFC volume susceptibility data at low temperatures indicates bulk superconductivity.}\label{zfc}
\end{figure}

The ZFC and FC temperature dependent magnetic susceptibility data are shown in Fig.~\ref{zfc}. Measurements were performed at 30 Oe in a temperature range from 1.8 K to 10 K. The onset of diamagnetism occurs near 6.6 K, which is in agreement with the transport data.  The ZFC data show a large, negative volume susceptibility of near $-$1 at low temperature.  A perfect Meissner fraction corresponds to $4\pi\chi = -1$.  Several small odd-shaped pieces of sample were measured, and the data were not corrected for demagnetization effects.

\subsection{Upper critical field}

\begin{figure}[h!]
\includegraphics[scale=0.45]{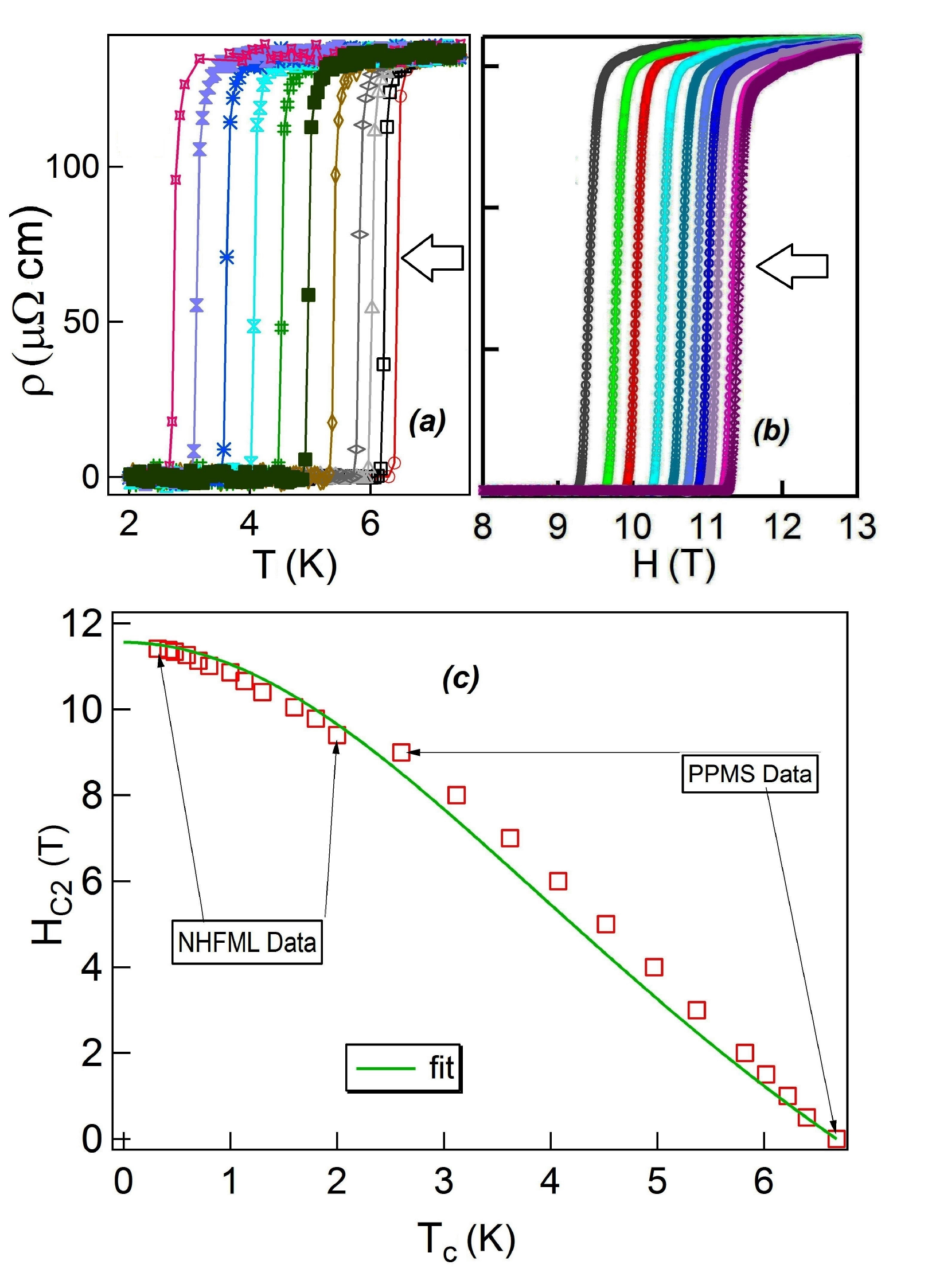}
\caption{Upper critical field of Re$_{6}$Zr as a function of temperature. The transition temperatures were taken from the midpoint of the resistivity drop from the normal state as seen in (a) and (b). In the left upper panel (a), the behavior of the transition temperature measured in PPMS under applied fields of 0, 0.5, 1, 1.5, 2, 3, 4, 5, 6, 7, 8, and 9 T is shown from right to left, respectively. On the right in plot (b), the critical fields at different temperatures down to 0.32 K were measured at NHFML. The solid line in (c) is a fit to Equation 3, as described in the text.}\label{upc}
\end{figure}

The upper critical field was calculated by applying a variety of magnetic fields up to 9 T to the same sample in the PPMS and measuring the shift in $T_{c}$. The transition temperatures at higher fields was measured in a 35-T resistive magnet at the NHFML. For these measurements, temperatures were as low as 0.32 K, and critical fields at different temperatures up to 2 K were investigated. At all applied fields, a sharp superconducting transition was observed as shown in the upper panel of Fig.~\ref{upc}.

As expected, $T_{c}$ shifts to lower temperature as the field increases and the transition gets broader. Superconductivity remains above 2 K for an applied field of 9 T, in the PPMS data, which indicates a large upper critical field and is consistent with data from NHFML. Using the midpoint resistivity from its normal state value, the upper critical field is plotted as a function of transition temperature, $T_{c}$. The variation of $H_{c2}$ with $T_{c}$ is approximately linear with a negative slope without showing any saturation for applied fields as high as 9 T. Below 2 K to 0.32 K and above 9 T, $H_{c2}$ shows slight saturation.

Using the Werthamer-Helfand-Hohenberg (WHH) expression~\cite*{w},
\begin{equation}
\mu_0 H_{c2} (T) = - 0.693\mu_0 (dH_{c2}/dT)_{T_c} T_c,
\end{equation}
$\mu_{0}$$H_{c2}$(0) was estimated using the data range from $T = T{_c}$ to  $T{_c}$/3. From the slope  $\mu_{0}$(d$H_{c2}$/dT)$\approx$ $\neg$ 2.31 $\pm$ 0.04 T/K, and using $T_{c}$ = 6.68 K, we found $\mu_{0}$$H_{c2}$(0) = 10.65 $\pm$ 0.02 T, which is slightly smaller than the value reported by Singh $et$ $al$\cite{sing}.

We also estimated the upper critical field by fitting the data with the empirical formula~\cite*{nb},
\begin{equation}
H_{c2}(T) = H_{c2}(0)(1-t^2)/(1+t^2),
\end{equation}
with $t$ = $T$/$T_{c}$, and $T_{c}$ the transition temperature at zero applied field. The fit gave a value of 11.6 $\pm$ 0.1 T. This value is close to the Pauli limiting field of 1.83$T_{c}$ , which is 12.22 T.
This large upper critical field can originate from the strong coupling or from a triplet pairing component in Re$_{6}$Zr, as well as a combination of the both.

If we assume the upper critical field to be purely orbital, the superconducting coherence length ($\xi$) can be calculated using $H_{c2}$(0) = $\Phi_{0}$/2$\pi$$\xi$$(0)^{2}$, where $\Phi_{0}$ = 2.0678 $\times$ $10^{9}$ Oe$\space$ \AA$^{2}$ is the flux quantum\cite{nb}. From this we found $\xi$(0) = 53.3 \AA, for $H_{c2}$(0) = 11.6 T.  Similarly, from the relation $H_{c1}$(0) = ($\Phi_{0}$/4$\pi$$\lambda^{2}$)$\ln$($\lambda$/$\xi$), using 8 mT as $H_{c1}$ which was reported in this ref~\cite*{singh}, the magnetic penetration depth was found to be, $\lambda$(0) = 3696 \AA. The Ginzburg-Landau parameter is then $k = \lambda/\xi$ = 69.3. The upper critical field and Pauli limiting field closely follow the relation $H_{c2}$(0)/$H_{\rm Pauli}$ = $\alpha$/$\sqrt{2}$. The Maki parameter was found to be $\alpha$ = 1.34. The sizable Maki parameter obtained is an indication that Pauli pair-breaking is non-negligible \cite{maki}. Thus, an anisotropic study of the upper critical field in a single crystal would provide greater evidence for it exceeding the Pauli limit, where the momentum-space dependence of the SO coupling could be studied\cite{prp2}. If it is found to be the case, the result would suggest a substantial contribution from a spin triplet component to the pairing mechanism\cite{book,mol}.

\subsection{Thermal conductivity}

\begin{figure}[h!]
\includegraphics[scale=0.40]{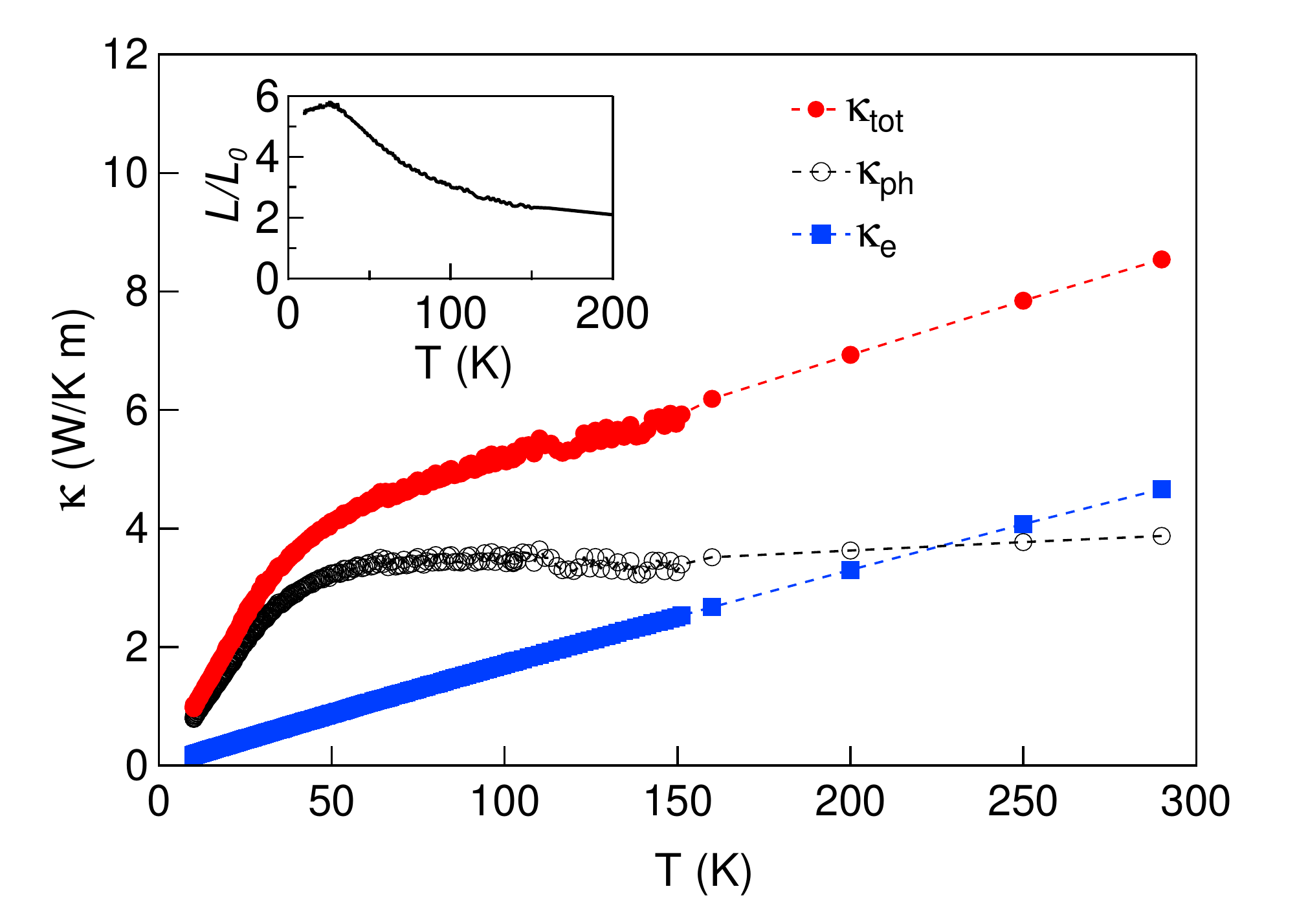}
\caption{Thermal conductivity of Re$_{6}$Zr above 10 K. The main panel shows the total, phonon, and electronic contributions to the thermal conductivity. The dashed lines are guides to the eye. Inset: Temperature dependence of the reduced Lorenz number.}\label{thermal}
\end{figure}

We measured thermal conductivity data between 2 - 290 K, with and without applied field. The zero-field data above 10 K are shown in Fig.~\ref{thermal}. A greater density of data points was taken below 150 K.  The total thermal conductivity has a room temperature value of about 8.5 W/K m, which is comparable to other metallic alloys.  We have assumed that the total thermal conductivity ($\kappa_{\rm{tot}}$) is composed of a lattice, or phonon contribution ($\kappa_{\rm{ph}}$), and a conduction electron contribution ($\kappa_{\rm{e}}$), which depends both on the temperature and carrier concentration. The electronic contribution to the thermal conductivity was estimated using the Wiedemann-Franz law, which assumes the energy/momentum relationship is given by a single parabolic band. Here, $\kappa_{\rm{e}}$ = $L_{0}T/\rho$, where $L_{0} = 2.45\times10^{-8}$ W $\Omega$ K$^{-2}$ is the Lorenz number. The phonon contribution was then estimated from $\kappa_{\rm{ph}}$ = $\kappa_{\rm{tot}}$$-$$\kappa_{\rm{e}}$. As shown in Fig.~\ref{thermal}, $\kappa_{\rm{e}}$ decreases proportionally with temperature, whereas $\kappa_{\rm{ph}}$ remains approximately constant above 50 K.

The temperature dependence of the reduced Lorenz number ($L/L_{0}$) is shown in the inset of Fig.~\ref{thermal}. Here, $L$=$\kappa_{\rm{tot}}$$\rho/T$. The reduced Lorenz number increases rapidly below 100 K with decreasing temperature and obtains a large maximum value of 5.8 at approximately 25 K.  These large values of $L/L_{0}$ are typically observed in heavy fermion compounds, such as URu$_2$Si$_2$ and CeCu$_4$Al~\cite*{behnia,falkowski}.  If the thermal conductivity was due solely to the electronic contribution, then the reduced Lorenz number would be identically 1, in accordance with the Wiedemann-Franz law.  Large values of $L/L_{0}$ suggest the thermal conductivity is dominated by phonons, especially below 100 K.

\begin{figure}[h!]
\includegraphics[scale=0.45]{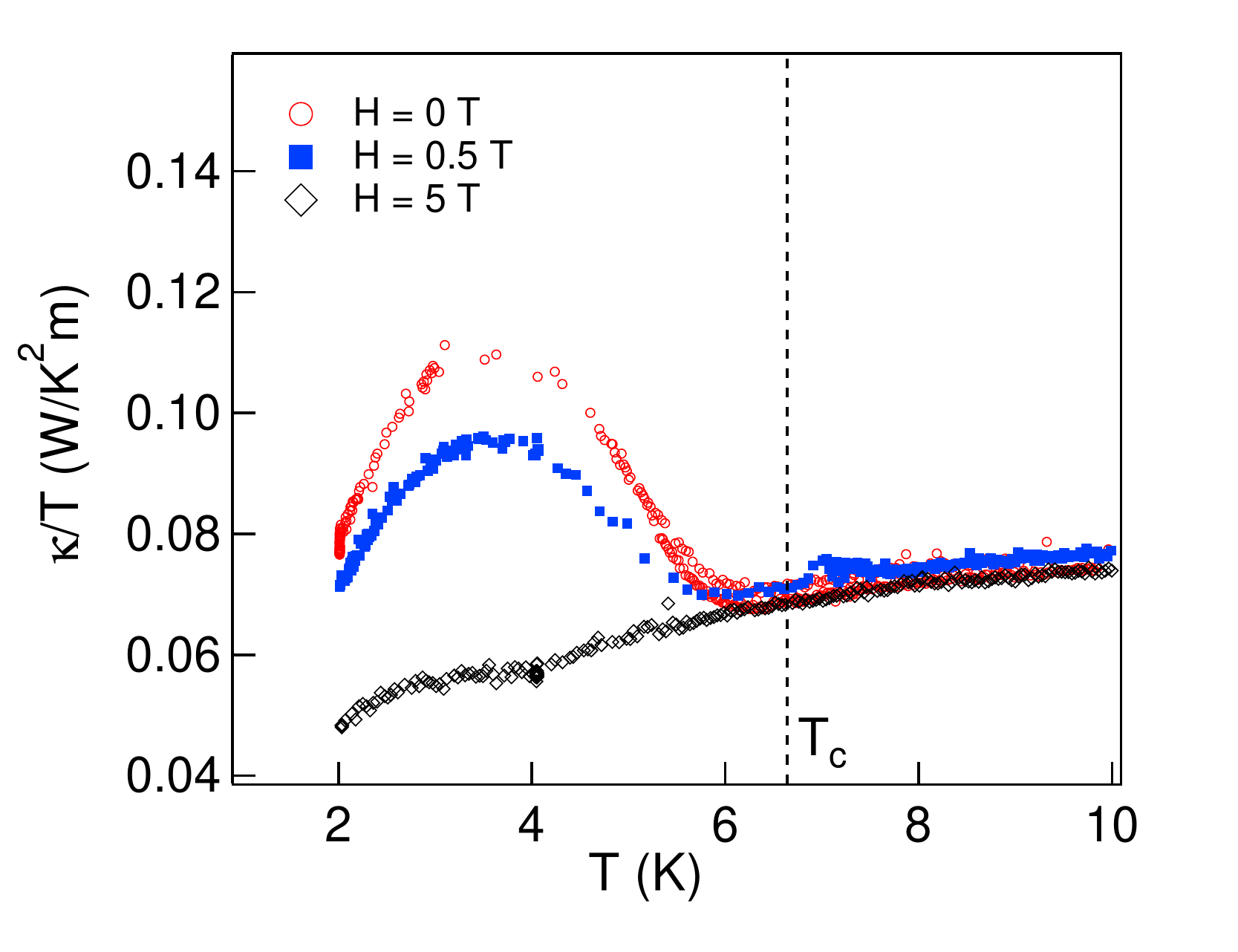}
\caption{Low temperature thermal conductivity of Re$_6$Zr in different magnetic fields, showing the enhancement below $T_{c}$ and the suppression with magnetic field.  The vertical dashed line indicates $T_{c}$ at zero field.}\label{themalenhance}
\end{figure}

Further evidence for a phonon-dominated thermal conductivity in Re$_{6}$Zr is shown in Fig.~\ref{themalenhance}, where we plotted the low temperature thermal conductivity below 10 K, which spans the superconducting transition.  The thermal conductivity shows no signature of superconductivity at the transition temperature, $T_{c}$, which is indicated by the dashed vertical line in Fig.~\ref{themalenhance}. However, a significant enhancement in $\kappa_{\rm{tot}}$ occurs below $T_{c}$. In zero field, the thermal conductivity increases below 6 K and reaches a maximum near 4.5 K. This behavior is typically observed in unconventional superconductors, such as heavy fermions, iron pnictides, and high-$T_{c}$ cuprates~\cite*{spin,behnia,zhang}, in contrast to conventional superconducting systems, where the thermal conductivity decreases below $T_{c}$ due to the loss of the electronic contribution\cite{bardeen}.

Many of the unconventional systems contain magnetic elements and order antiferromagnetically. The enhancement in the thermal conductivity is often attributed to a reduction in scattering from spin fluctuations~\cite*{spin}. We do not expect this to be the mechanism responsible for the enhancement in the thermal conductivity in Re$_6$Zr, as it displays Pauli paramagnetism below room temperature.

The enhancement could be electronic in origin, due to strong inelastic scattering of electrons that freezes out with the opening of the gap. It could also be attributed to the phonon component, as this contribution will increase below $T_{c}$ due to a rapid decrease in the quasiparticle scattering. However, a reasonable fit of the resistivity to the Fermi-liquid $T^2$ dependence, a small RRR value, and the estimated dominance of the phonon contribution to $\kappa$ at the superconducting transition are in favor of the latter scenario. As shown in Fig.~\ref{themalenhance}, the peak in the thermal conductivity is suppressed in the presence of a magnetic field. We interpret the suppression as due to phonon-vortex scattering in the mixed state ($H_{c1}$$\textless$$H$$\textless$$H_{c2}$), where $H_{c1}\approx$ 8 mT~\cite*{singh}. This requires the intervortex spacing to be less than the phonon mean free path.  Assuming a triangular vortex lattice, the vortex spacing ($a_v$) can be estimated by~\cite*{tinkham} $a_v=1.075(\Phi_{0}/H)^{1/2}\approx700$ \AA, for a field value of 0.5 T. From specific heat and thermal conductivity data, and assuming a reasonable sound velocity of 3500 m/s, the phonon mean-free path is estimated to be $\approx$ 3000 \AA, which is indeed larger than the vortex spacing, further supporting the notion of the thermal conductivity being dominated by the phonon contribution in zero field.

\begin{figure}[h!]
\includegraphics[scale=0.42]{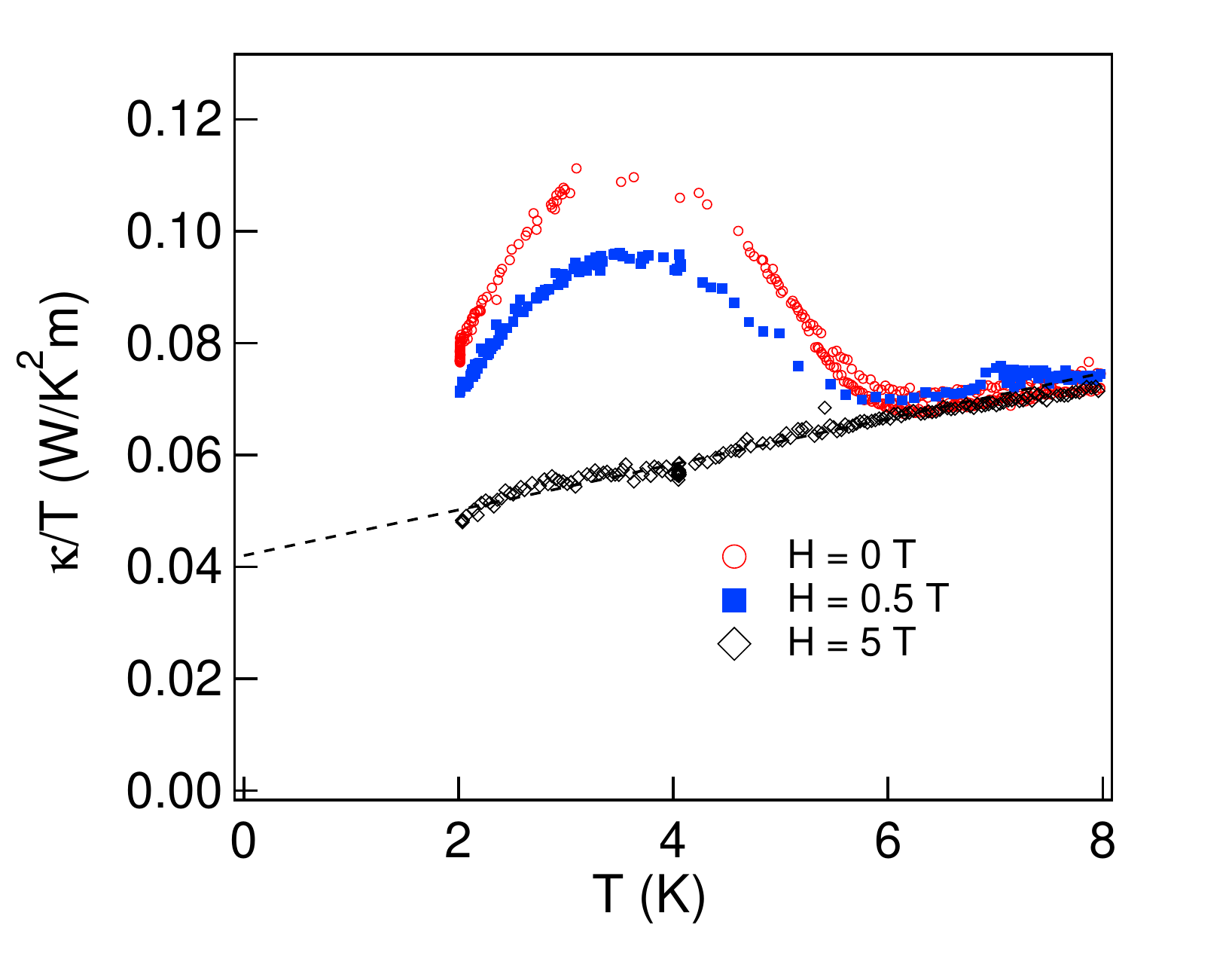}
\caption{Temperature dependence of thermal conductivity $\kappa$$_{tot}$ below $T_{c}$. The dashed line is a linear fit to the data.}\label{thermallowt}
\end{figure}

The temperature dependence of $\kappa/T$ is shown in Fig.~\ref{thermallowt}. Linear fit (dashed line) match the data well below the phonon enhancement peak for zero field and 0.5 T, and to even higher temperatures for the 5 T data, in which the phonon peak has been completely suppressed. A similar linear temperature dependence in $\kappa/T$ was observed in CePt$_{3}$Si~\cite*{cekt}. While it may be tempting to draw conclusions about the existence of nodes from the extrapolated linear dependence to $T=0$, this would be misleading, as detailed measurements at much lower temperatures are necessary to determine the behavior of the electronic component of $\kappa/T$.

\subsection{London penetration depth}

\begin{figure}[h!]
\includegraphics[scale=0.45]{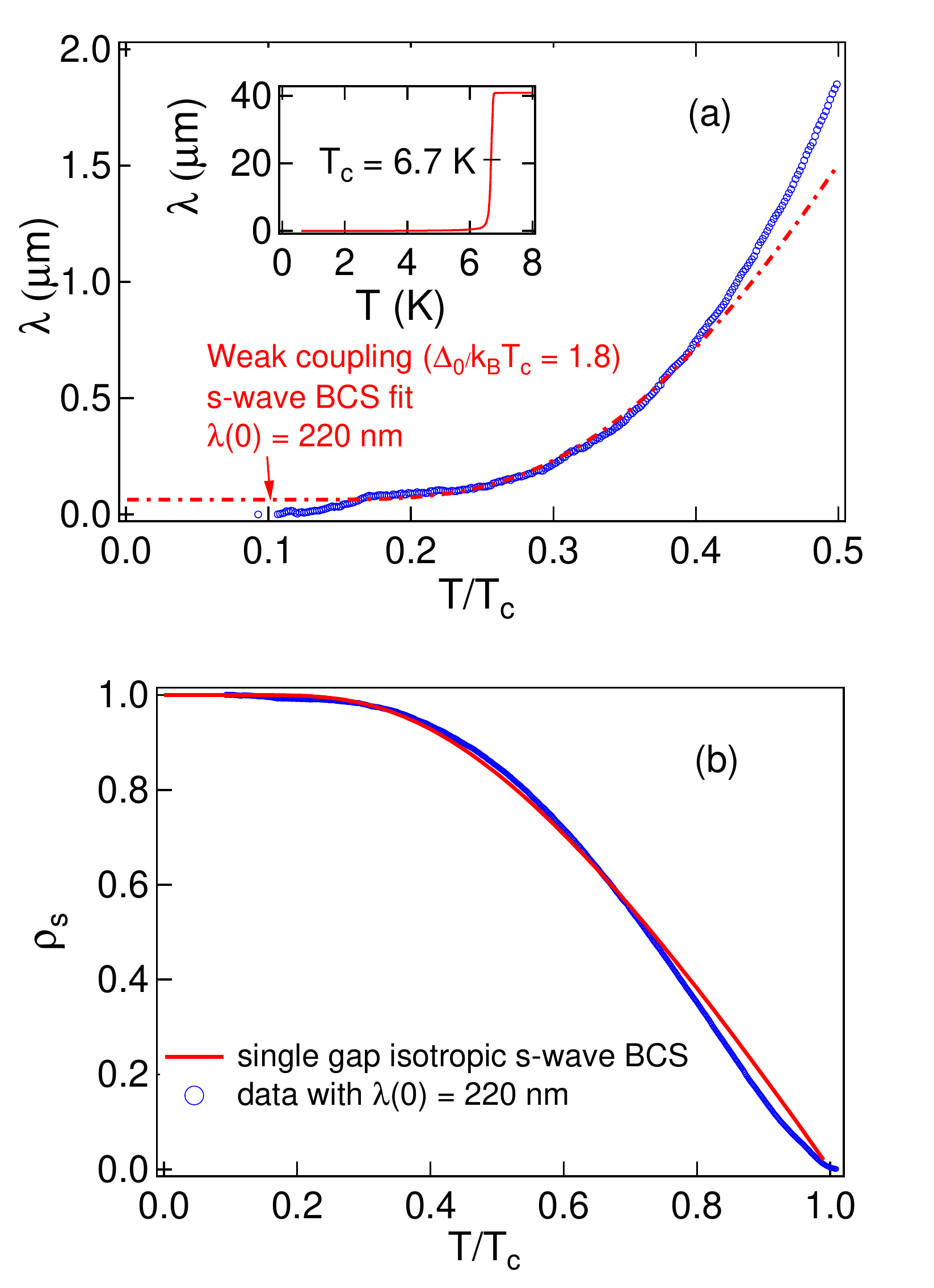}
\caption{Temperature dependence of the London penetration depth (a) and superfluid density (b). The superconducting transition at $T_{c}$ = 6.7 K is shown in the inset of the upper panel. The dotted line in the main body of the upper panel is a BCS fit to the data below $T = T_{c}$/3 as described in the text. The lower panel shows the corresponding superfluid density, and the solid line is a BCS fit to the data as described in the text.}\label{penetration}
\end{figure}
The temperature dependence of the London penetration depth is shown in Fig.~\ref{penetration}. A sharp superconducting transition was observed at 6.7 K, which is similar to the value from the resistivity data. In order to determine the structure of the gap, one needs to probe the behavior of $\Delta$$\lambda$($T$) below about $T_{c}$/3 (2.23 K in the current study). We measured $\Delta$$\lambda$($T$) down to 0.5 K ($\sim$0.07 $T_{c}$) as seen in Fig.~\ref{penetration}(a). While a small impurity feature was noticed below 1.1 K (0.16 $T_{c}$), we were able to observe the clear temperature-independent behavior of $\Delta$$\lambda$($T$) below 0.25 $T_{c}$ which is consistent with the s-wave BCS fit of

\begin{equation}
\Delta\lambda(T) \approx\Delta\lambda(0)\sqrt{\pi\Delta(0)/2k_{B}T)} exp(-\Delta(0)/ k_{B}T)
\end{equation}

with $\Delta$$\lambda$(0)=220 nm. This suggests an isotropic superconducting energy gap similar to Mo$_{3}$Al$_{2}$C~\cite*{depth4}. Interestingly, other physical and thermal property studies of Mo$_{3}$Al$_{2}$C, such as power law behavior in the NMR relaxation rate and absence of a Hebbel-Slichter peak,  deviate from the BCS prediction, suggesting the possibility of a nodal superconducting gap in Mo$_{3}$Al$_{2}$C~\cite*{mol,moli2}.  In addition, Fig.~\ref{penetration}(b) shows the corresponding superfluid density of Re$_6$Zr,  $\rho$$_{\rm s}$($T$) = ($\Delta$$\lambda$(0)/$\Delta$$\lambda$($T$))$^{2}$  which is also consistent with that of a single s-wave isotropic BCS gap. In all likelihood, the polycrystalline nature of our sample might prevent us from determining its actual gap structure due to disorder and or impurities, as suggested by the large residual resistivity and small RRR value.

\subsection{Heat capacity}

A characteristic superconducting transition was observed in the specific heat data, indicating bulk superconductivity. By fitting the $C/T$ vs $T^{2}$ data to the equation $C/T$ = $\gamma$+$\beta$$T^{2}$, the value of the Sommerfeld coefficient was determined to be $\gamma$ = 27.5 $\pm$ 0.4 mJ mol$^{-1}$K$^{-2}$ and $\beta$ = 0.451 $\pm$ 0.009 mJ mol$^{-1}$K$^{-4}$. The ratio $\Delta$$C$/$\gamma$$T_{c}$ was found to be 1.62, which is in agreement with the previous work\cite{sing}. The ratio $\Delta$$C$/$\gamma$$T_{c}$ is larger than the BCS value of 1.43, indicating moderate coupling strength in Re$_6$Zr, which is in agreement with the behavior of the thermal conductivity.

\begin{figure}[h!]
\includegraphics[scale=0.40]{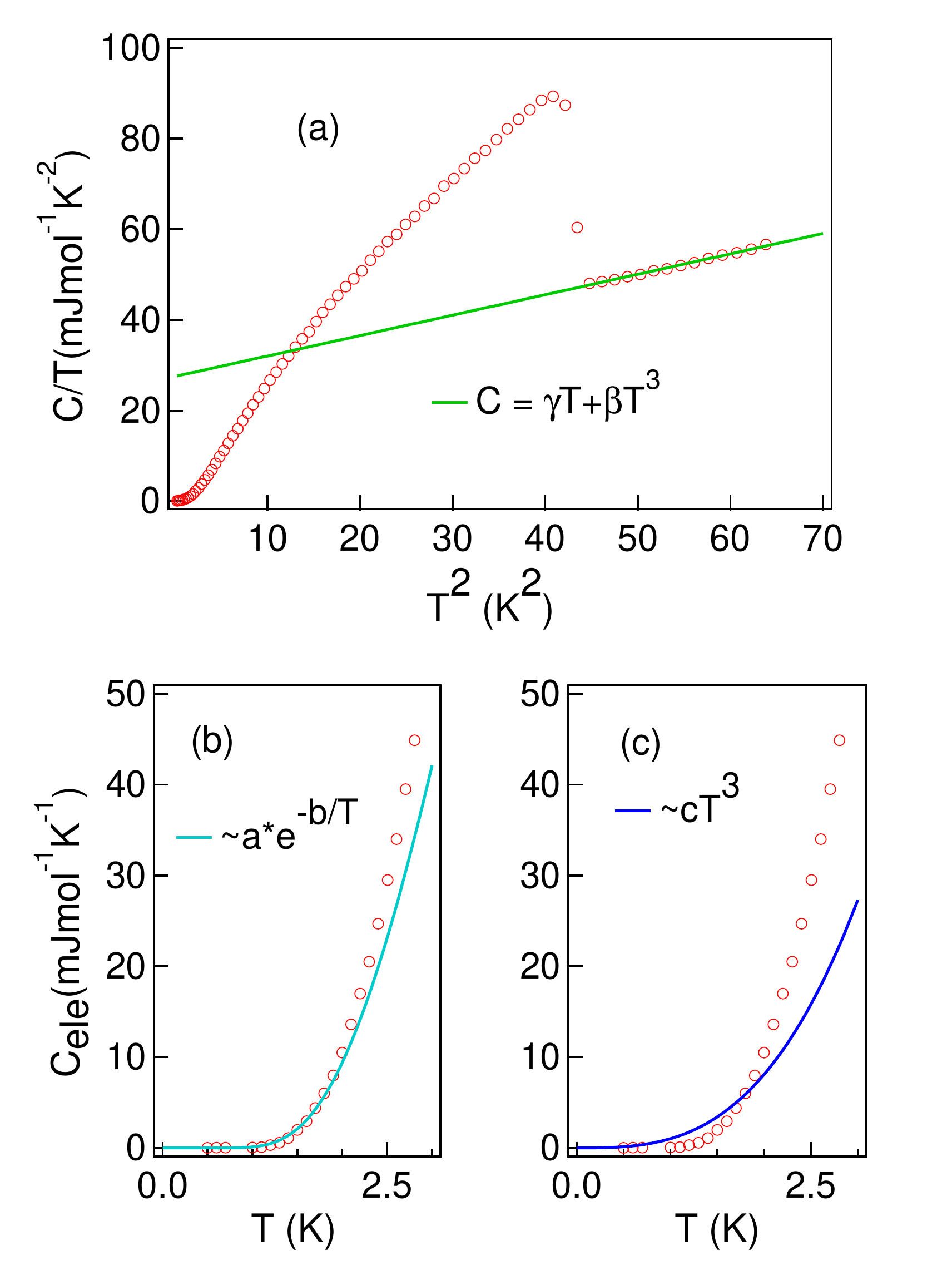}
\caption{Temperature dependence of the total and electronic specific heat. Upper panel (a) shows the superconducting transition and estimation of the phonon and electronic contribution using the fit described in the text(solid lines). The electronic specific heat below $T_{c}$ is shown in the lower two panels with fits for the BCS (b) and anisotropic (c) gap structures (solid lines).}\label{heatcapacity}
\end{figure}

The low temperature behavior of the electronic specific heat and thermal conductivity is often a good indicator of the superconducting gap characteristics. For example, a power law behavior indicates nodes in the superconducting energy gap, while exponential behavior indicates a conventional, fully-gapped BCS state~\cite*{book}. The electronic specific heat ($C_{\rm ele}$) below $T_{c}$ was estimated by subtracting the lattice component $\beta$$T^{3}$, from the total specific heat. The electronic specific heat was then analyzed by fitting the data to the following forms:  $a$ $\exp^{-b/T}$  and  $c$$T^{3}$. These are the expected temperature dependencies for gaps which are isotropic or contain point nodes, respectively. Our data are well fit by the exponential fit $C_{\rm ele}$ $\propto$ $a$ $\exp^{-b/T}$ below $T$ = 2 K, as shown in Fig.~\ref{heatcapacity}(b). The cubic power law does not represent the data as well as the exponential fit [Fig.~\ref{heatcapacity}(c)]. The above analysis suggests an isotropic gap. However, the accuracy of low-temperature electronic specific-heat data obtained by subtracting a phonon contribution grossly depends on the accuracy in determining the normal-state heat capacity from the in-field measurements\cite{moli2}. In our study, the phonon contribution was estimated from zero field data. Hence, the actual temperature dependence of the electronic specific heat might not be represented by the current data. Therefore, it would be difficult to distinguish the power law due to point nodes from the exponential behavior in our measurements. Note also that a similar behavior of the low temperature electronic specific heat in Mo$_{3}$Al$_{2}$C was found in two different studies. First, Karki $et$ $al$~\cite*{mol} found exponential behavior in the low temperature electronic specific heat by using zero field data to calculate the phonon contribution, while Bauer $et$ $al$~\cite*{moli2} found a power law behavior when the phonon contribution was calculated by investigating the heat capacity thoroughly in both zero and applied field.

\subsection{Doping and pressure studies}

\begin{figure}[h!]
\includegraphics[scale=0.40]{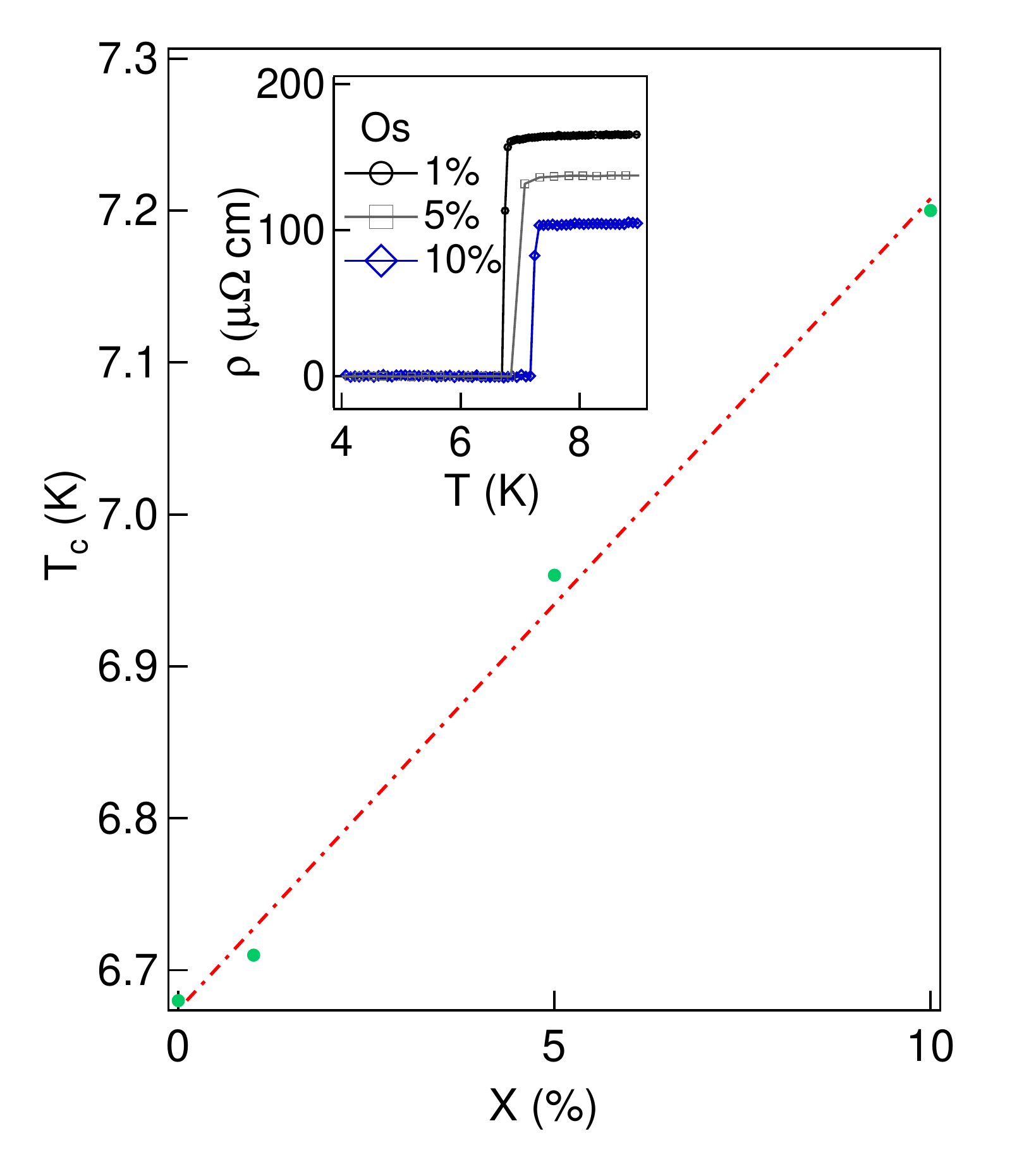}
\caption{Change in the superconducting transition temperature as a function of Os doping. The broken line is a linear fit to the data. Inset: Superconducting transitions in the Os-doped samples observed in the temperature dependence of the electrical resistivity.}\label{osdoping}
\end{figure}

To investigate the effect of chemical doping on the superconducting transition temperature of Re$_6$Zr, we synthesized multiple samples doped with Hf, Ti, W and Os. Among the doped samples, Os doping had a significant positive effect on the transition, enhancing $T_{c}$ by $\approx$ 1\% per 1\% doping concentration as seen in Fig.~\ref{osdoping}. Due to the heavy nature of its constituent elements, we expect SO coupling to play a significant role in the physical properties of Re$_6$Zr. Os, being heavier, could enhance the SO coupling leading to an increase in $T_{c}$.  The other dopants at or below 10\% lowered $T_{c}$ or had little effect. Note that unconventional pairing mechanisms in anisotropic channels are sensitive to disorder, and doping generally suppresses the transition temperature. An increase in $T_c$ with Os doping then either indicates an $s$-wave singlet-triplet mixing, inconsistent with the TRS breaking, or an intricate interplay of the ASOC with the microscopic mechanism responsible for unconventional pairing. The latter possibility, to our knowledge, has not yet been investigated theoretically in sufficient detail.

\begin{figure}[h!]
\includegraphics[scale=0.50]{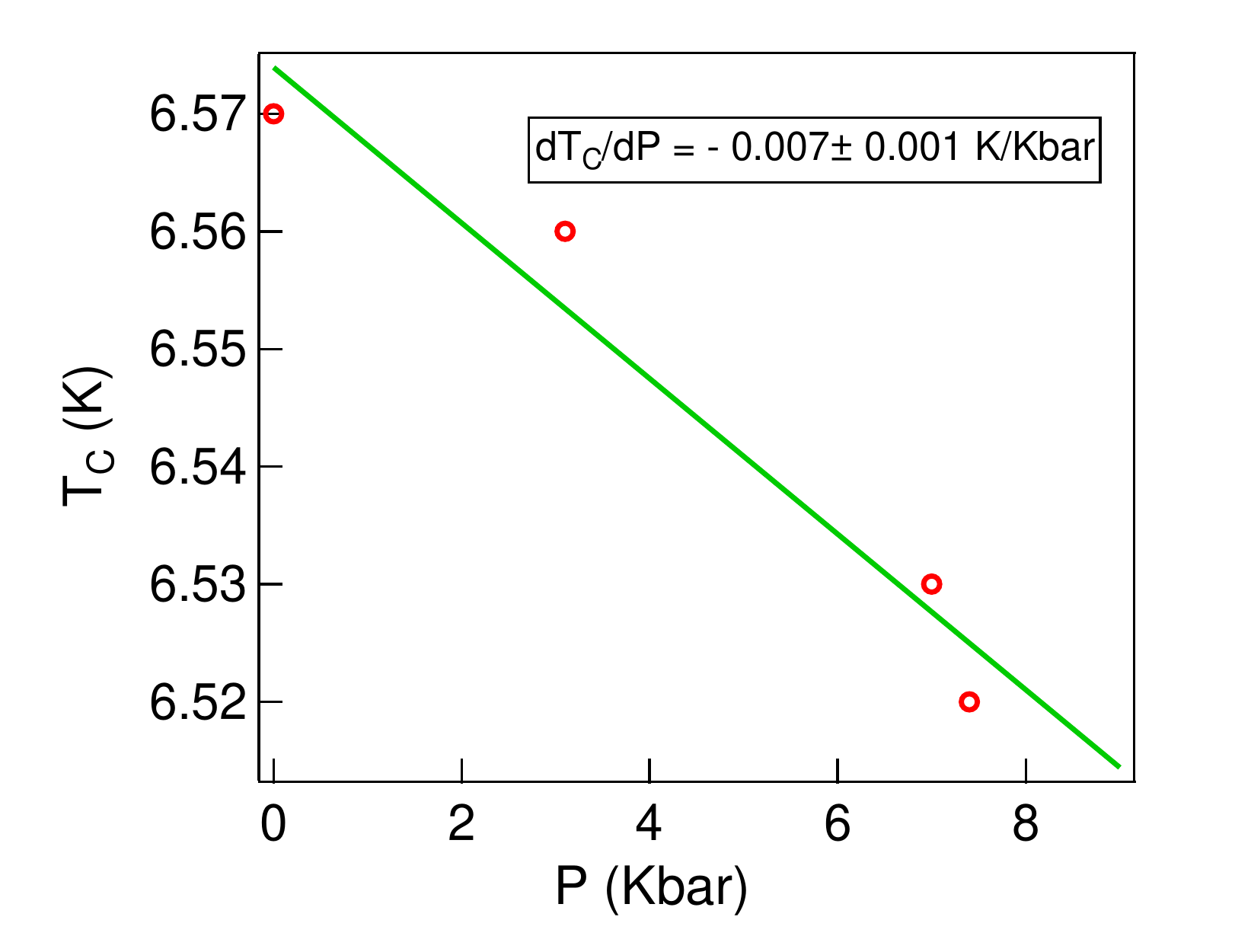}
\caption{Change in the superconducting transition temperature of pure Re$_6$Zr due to applied pressure. Solid line is a linear fit to data which indicates a negative slope.}\label{pressure}
\end{figure}

The pure sample was also exposed to moderate hydrostatic pressure up to 8 kbar. Pressure only had a small effect on the transition, decreasing $T_{c}$ slightly as shown in Fig.~\ref{pressure}. The solid line is a linear fit to data which shows the decreasing trend of transition temperature under applied pressure\\

\section{conclusion}

Polycrystalline samples of pure and doped Re$_6$Zr were prepared by arc-melting techniques. The results of XRD measurements and elemental analysis confirm single phase materials with the noncentrosymmetric $\alpha$-Mn structure type. From resistivity, magnetic susceptibility, and specific heat measurements, Re$_6$Zr was confirmed to be a strongly correlated, type-II superconductor with a bulk transition temperature near 6.7 K with a strong electron-phonon coupling. Several doping studies revealed that Os doping significantly enhanced $T_{c}$, which is possibly due to enhanced SO coupling. The thermal conductivity is dominated by the phonon contribution near the superconducting transition and is enhanced below $T_{c}$ due to a decrease in electron-phonon scattering, as normal electrons pair up to form the condensate. Applied magnetic fields between $H_{c1}$ and $H_{c2}$ suppress the peak due to phonon-vortex scattering. The upper critical field $H_{c2}$(0) is comparable to the calculated Pauli limit, which can be a consequence of the SO interaction or strong coupling or might also be due to the contribution from a triplet pairing component to the order parameter. While the above measurements suggest that the superconducting behavior of Re$_{6}$Zr deviates from that of the conventional superconductors, the low temperature electronic specific heat and penetration depth are best fit with an exponential temperature dependence. The data, taken in its entirety, somewhat weigh against lines of nodes in the superconducting gap, but may be consistent with point nodes. To truly establish the effect of broken inversion symmetry in Re$_{6}$Zr, low temperature studies of the physical and thermal properties on a single crystal are highly desirable.

\acknowledgments
We acknowledge useful communications with J. Quintanilla and J. F. Annett. We also acknowledge Dr. Clayton Loehn at Shared Instrumentation Facility (SIF), LSU for chemical analysis. D.P.Y. acknowledges support from the NSF under Grant No. DMR-1306392.  P.W.A. acknowledges support by the U.S. Department of Energy (DOE) under Grant No. DE-FG02-07ER46420, I.V. acknowledges support from NSF Grant No. DMR-1410741, and S.S. also acknowledges the DOE Office of Science, Basic Energy Sciences (BES), under Award No. DE-FG02-13ER46946. A portion of this work was performed at the National High Magnetic Field Laboratory, which is supported by National Science Foundation Cooperative Agreement No. DMR-1157490 and the State of Florida.Work in Ames was supported by the U.S. DOE, Office of Science, Basic Energy Sciences, Materials Science and Engineering Division. Ames Laboratory is operated for the U.S. DOE by Iowa State University under contract DE-AC02-07CH11358.

\bibliography{mybib2}

\end{document}